\documentclass[%
 reprint,
nofootinbib,
 amsmath,amssymb,
 aps,
]{revtex4-2}

\usepackage{graphicx}
\usepackage{dcolumn}
\usepackage{bm}

\usepackage{enumerate}

\begin{document}

\preprint{APS/123-QED}

\title{Ghost projection via focal-field diffraction catastrophes}

\author{James A. Monro}
\author{Andrew M. Kingston}%
 \altaffiliation[Also at ]{CTLab: National Laboratory for Micro Computed-Tomography, Advanced Imaging Precinct, The Australian National University, Canberra, ACT 2601, Australia.}
\affiliation{%
 Department of Materials Physics, Research School of Physics,\\
 The Australian National University, Canberra ACT 2601, Australia
}%
\author{David M. Paganin}
 \email{david.paganin@monash.edu}
\affiliation{
School of Physics and Astronomy,
Monash University, Victoria 3800, Australia
}

\date{\today}

\begin{abstract}
Ghost projection is the reversed process of computational classical ghost imaging that allows any desired image to be synthesized using a linear combination of illuminating patterns. Typically, physical attenuating masks are used to produce these illuminating patterns.  A mask-free alternative form of ghost projection is explored here, where the illuminations are a set of caustic-laden diffraction patterns known as diffraction catastrophes. These are generated by focusing a coherent beam with spatially modulated phase having random Zernike-polynomial aberrations. We demonstrate, via simulation, that a suitable linear combination of such random focal-field intensity patterns can be used as a basis to synthesize arbitrary images.  In our proof-of-concept ghost-projection synthesis, the positive weighting coefficients in the decomposition are proportional to exposure times for each focal-field diffraction catastrophe. Potential applications include dynamic on-demand beam shaping of focused fields, aberration correction and lithography.
\end{abstract}

\maketitle


\section{Introduction}

Ghost projection is a method for producing any desired spatial distribution of radiant exposure by superimposing a set of pre-determined illuminating patterns \cite{paganin2019writing,ceddia2022ghost,ceddia2022ghostII,ceddia2023universal,kingston2025neutron}. In essence, the required image is decomposed into a multiple exposure of fixed random patterns. The patterns are assigned different positive weights (e.g.~exposure times) which depend on the desired image.  Ghost projection may be viewed as the reverse process to that of classical computational ghost imaging \cite{shapiro2008computational}, from which it inherits its name.

To the best of our knowledge, at the time of writing there are five papers on ghost projection \cite{paganin2019writing, ceddia2022ghost, ceddia2022ghostII, ceddia2023universal,kingston2025neutron}.  All of these publications employ a mask-based strategy whereby a variety of spatially-random attenuating masks are illuminated to generate a set of spatially-random patterns, which can approximate a basis for the space of image projections. 
However, at small detail scales, attenuating masks suffer from diffraction effects that distort the patterns, impeding their effectiveness in ghost projection. 

\begin{figure}[!htb]
    \centering
    \includegraphics[width=\linewidth]{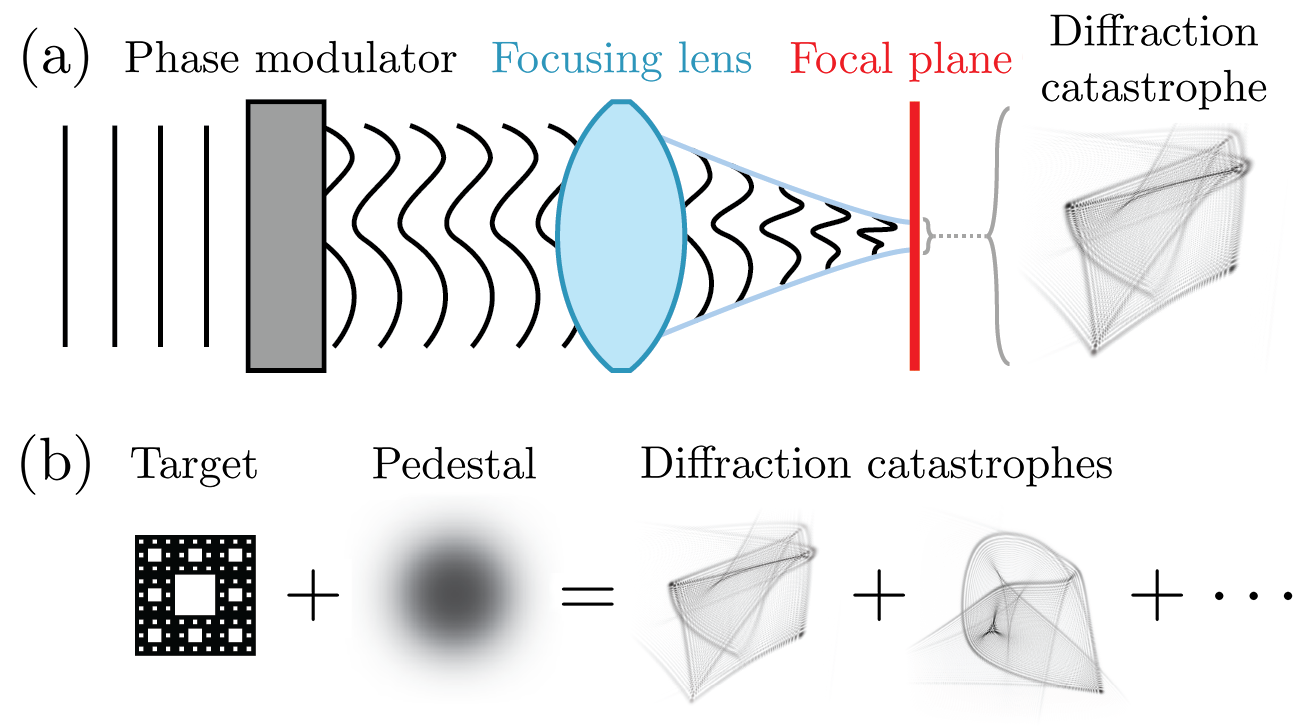 }
    \caption{Schematic of diffraction-catastrophe ghost projection. (a) Parallel lines on the left represent a coherent plane wave traveling to the right, with curved wavefronts indicating the wavefield phase has been modulated (aberrated). The transverse intensity pattern at the focal plane is the ``diffraction catastrophe'' \cite{kravtsov2012caustics}. (b) Superposing many different focal-plane diffraction catastrophes, for a variety of modulations, is the essence of our technique for ``ghost projection'' synthesis of an arbitrary target image \cite{paganin2019writing,ceddia2022ghost,ceddia2022ghostII,ceddia2023universal,kingston2025neutron}. Note that the ``pedestal'' is a smooth background that is well approximated by the ensemble average of the utilized focal-field catastrophes.}
    \label{fig:Generating diffraction catastrophes}
\end{figure}

In this paper, we propose a pattern generation procedure that forgoes attenuating masks and \textit{exploits} diffraction, rather than being limited by it. In this context, consider a coherent collimated beam that is incident upon a focusing lens, thereby producing an Airy disk at the focal plane \cite{born2013principles}. If the phase of the beam is modulated non-uniformly before reaching the lens (so that a cross section of the beam has smoothly-varying non-constant phase), then we generally observe a highly-structured diffraction pattern (``diffraction catastrophe''\footnote{Here and henceforth, the term ``catastrophe'' is used in the technical sense of ``catastrophe theory'' \cite{PostonStewartTheoryBook}.  In particular, ``diffraction catastrophes'' \cite{kravtsov2012caustics} refer to the caustic lines of the geometrical-optics (zero-wavelength) limit, which are softened by the nonzero wavelength so as to be ``decorated'' with fine diffraction features (Ref.~\cite{Nye1999}, p.~2).}) at the focal plane instead of an Airy disk (see Fig.~1(a)). The envisioned advantage of this method is that the phase function may be undetailed and slowly varying (aiding ease of fabrication) and yet produce a highly detailed diffraction pattern. 
It is natural to ask whether an ensemble of such diffraction catastrophes---produced by an ensemble of different phase modulations that are sufficiently strong to produce caustics---may be used as a basis for the purposes of ghost projection. This key idea, suggested in Ceddia {\em et al.}~\cite{ceddia2023universal}, is explored in the present paper.

In addition to our blue-sky interest in focal-field ghost projection, three potential future practical applications motivate the present work: 

\begin{enumerate}[(i)]

\item For systems in which many focal-field diffraction catastrophes may be sequentially generated in a reproducible manner, the method provides an avenue for the {\em dynamic on-demand beam shaping of focused fields}. A dynamic sequence of on-demand intensity patterns (or frames), $\{F_1,F_2,...\}$, with a given frame rate, can be created by these focal-field diffraction catastrophes. Assuming that approximately $\mathcal{N}$ focal-field intensity patterns, $\{I_1(\mathbf{r}), I_2(\mathbf{r}),\cdots\,I_{\mathcal{N}}(\mathbf{r})\}$, can be generated in the time of one frame, then each on-demand intensity pattern in the sequence, $F_t$, can be generated via the linear combination
\begin{equation}
F_t(\mathbf{r}) := \sum_{k=1}^{\mathcal{N}}\tau_k I_k(\mathbf{r}),     
\end{equation}
where $\tau_k$ represents the associated exposure time\footnote{Stated more precisely, $\tau_k$ is a dimensionless non-negative weighting coefficient that is proportional to an exposure time.} of each illuminating pattern, $I_k$. Here and henceforth, $\mathbf{r}$ denotes transverse coordinates in the focal plane (see Fig.~\ref{fig:Generating diffraction catastrophes}(a)).

\item Ghost projection using diffraction catastrophes {\em circumvents the issue of imperfect focusing elements}: so long as we include these imperfections when calculating our diffraction catastrophes, the weight-generating algorithm will naturally generate a different set of exposure times to compensate, and the image reconstruction quality will be unaffected. This requires either that the aberrations (imperfections) are sufficiently well-characterized such that the simulated diffraction catastrophes are accurate, or that the diffraction catastrophes are measured empirically. For instance, if one desires to project a single spot, this form of ghost projection can be used as an indirect form of aberration correction, producing a non-distorted spot at the cost of the ``pedestal'' mentioned in the caption of Fig.~\ref{fig:Generating diffraction catastrophes}. 

\item Given the capacity for miniaturization inherent in a focused-beam geometry, {\em the method might also be useful for lithography}, e.g.~using aberrated electron-beam foci \cite{ReimerKohlBook} or aberrated x-ray-beam foci \cite{Paganin2006, JizhouLi2023}.  

\end{enumerate}
While the specifics of the preceding possible future applications lie beyond the scope of the present paper, they provide important context which guides many of the ideas explored here.

As additional context, it is clarifying to recall that classical ghost imaging is a computational imaging technique where a set of known illuminating patterns are incident upon an unknown sample. For each illumination pattern, the corresponding measurements of total intensity transmitted (or reflected) by the sample, referred to as ``bucket values,'' are recorded by a single-pixel sensor. Neither the set of intensity patterns nor the bucket values alone enable one to determine any positional information about the unknown sample. However, an image of the sample can be reconstructed by exploiting the correlation of the two sets of measurements, i.e., pattern and bucket value pairs \cite{katz2009compressive, bromberg2009ghost, erkmen2010ghost, padgett2017introduction}. Conversely, in ghost projection \cite{paganin2019writing,ceddia2022ghost,ceddia2022ghostII,ceddia2023universal,kingston2025neutron}, the patterns and desired image are known but the required exposure times/weights (analogous to positive bucket values) are unknown, and must be found.

In this simulation-based paper we explore the efficacy of diffraction catastrophes as an illuminating-pattern basis for ghost projection of different target images. We compare algorithms for generating coefficients of these patterns to produce a desired image. We also examine the properties of pedestals (additive offsets) in images generated by this set. To develop these key themes, the remainder of the paper is structured as follows. Section \ref{sec:Background} outlines the background theory and approach regarding ghost projection.  Section~\ref{sec:DiffCatastrophesAsRandomBasis} extends this to the case of using diffraction catastrophes as a random basis for the purposes of ghost projection. The modeling of focal-field diffraction catastrophes is outlined in Sec.~\ref{sec:MethodsMainHeading}. Section \ref{sec:Results and analysis} contains the simulated ghost-projection results, followed by a discussion in Sec.~\ref{sec:Discussion}.  Some practical considerations and possible avenues for future work are outlined in Sec.~\ref{sec:PracticalConsiderationsAndFuturework}. Concluding remarks are given in Sec.~\ref{sec:Conclusion}.

\section{Principles of ghost projection using a speckle-field basis}
\label{sec:Background}

\begin{figure}[!htb]
    \centering
    \includegraphics[width=\linewidth]{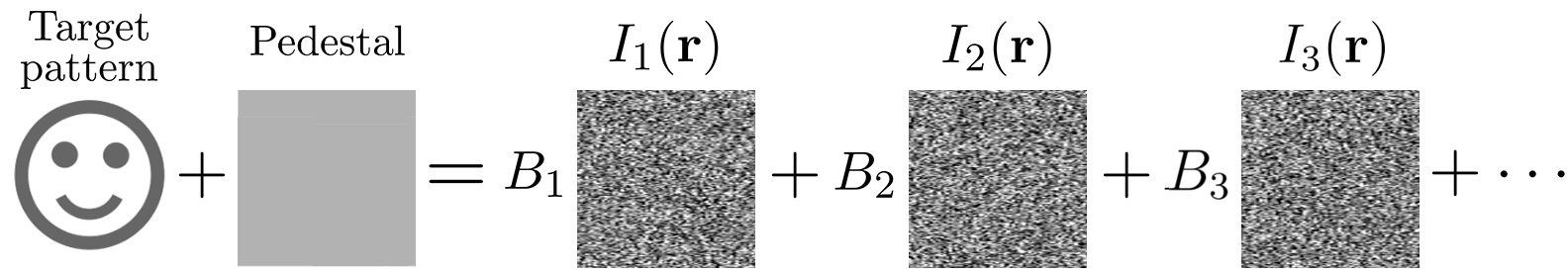}
    \caption{Illustration of the ghost-projection concept, for the case of spatially-random speckle maps, in the form given by Eq.~(\ref{eq:VanillaGhostProjection2}). Here, the smoothed target pattern (the smiling face) is $S(\mathbf{r})\otimes PSF(\mathbf{r})$, the pedestal (or offset) is $\langle B_k\rangle \overline{I}$, and each speckle pattern $ I_k(\mathbf{r})$ is weighted by the positive coefficient $B_k$.}
    \label{fig:VanillaGhostProjection}
\end{figure}

In this section we outline the principles of ghost projection as established in \citet{paganin2019writing} and Ceddia {\em et al.}~\cite{ceddia2022ghost, ceddia2022ghostII}. Let $\{I_k(\mathbf{r})\}_{k \in \{1,\cdots,\mathcal{N}\}}$ define a set of $\mathcal{N}$ spatially-random intensity images, such as a statistical ensemble
\begin{equation}
\{I_1(\mathbf{r}),I_2(\mathbf{r}),\cdots, I_\mathcal{N}(\mathbf{r})\}:=\{I_k(\mathbf{r})\}, \quad k=1,2,\cdots,\mathcal{N}
\end{equation}
of speckle maps \cite{GoodmanSpeckleBook} that each have the same correlation length\footnote{Here, the ``correlation length,'' $\ell$, of the speckle map refers to the width of the autocovariance function, for the intensity distribution associated with any particular speckle map.  The assumption of statistical spatial stationarity implies $\ell$ is independent of position, for any one speckle map.  We also implicitly assume ergodicity, so that spatial averages over one speckle map may be replaced with ensemble averages over the set of all such speckle maps, considered as a stochastic process. } $\ell$ which is independent of position. As suggested by the sketch in Fig.~\ref{fig:VanillaGhostProjection}, for sufficiently large $\mathcal{N}$ this set has many of the characteristics that one would expect of a complete basis. In particular, if we let $\overline{I_k} \in \mathbb{R}^+$ denote the spatially-averaged intensity of the $k\text{th}$ intensity pattern, then the set $\{I_k(\mathbf{r})-\overline{I_k}\}$ will be near-orthogonal (in the sense that the inner product of any two different patterns is expected to be small relative to the inner product of a pattern with itself). Accordingly, such a set can be employed for the synthesis of functions. 

To further develop this latter point, recall the completeness relation associated with a genuinely-complete infinitely-large set of real functions $\{\psi_k(\mathbf{r})\}$ \cite{Messiah}, namely  
\begin{equation}\label{eq:completenessRelation}
\langle \psi_k(\mathbf{r}) \psi_k(\mathbf{r}')\rangle  = \delta(\mathbf{r}-\mathbf{r}').
\end{equation}
Here the angular brackets denote ensemble average over $k \in \{1,\cdots,\mathcal{N}\}$ and $\delta(\mathbf{r})$ is the Dirac delta. In our case, rather than the preceding ``ideal'' completeness relation, we instead have the smoothed completeness relation \cite{ferri2010differential, PellicciaIUCrJ, paganin2019writing}
\begin{equation}\label{eq:smoothedCompletenessRelation}
\langle [I_k(\mathbf{r})-\overline{I_k}][I_k(\mathbf{r'})-\overline{I_k}]\rangle = \Upsilon~PSF(\mathbf{r}-\mathbf{r}').
\end{equation}
Here, $PSF(\mathbf{r})$ is a rotationally symmetric point-spread function with characteristic width $\ell$, and $\Upsilon$ is a normalization constant that enforces the demand that the area under the point-spread function be equal to unity. From a stochastic perspective, Eq.~(\ref{eq:smoothedCompletenessRelation}) states that the autocovariance of the background-subtracted speckle-field intensity decays to zero over a spatial scale on the order of the speckle size.  

From an image-synthesis perspective that encapsulates the ghost-projection concept, let $S(\mathbf{r})$ be an image that one wishes to synthesize.  Upon multiplying both sides of Eq.~(\ref{eq:smoothedCompletenessRelation}) by $S(\mathbf{r}')$, integrating over $\mathbf{r}'$ and using $\otimes$ to denote two-dimensional convolution, we obtain
%
%
\begin{equation}\label{eq:VanillaGhostProjection}
S(\mathbf{r})\otimes PSF(\mathbf{r}) = \langle (B_k - \overline{I_k} \:\overline{S\:}) [I_k(\mathbf{r})-\overline{I_k}]\rangle.
\end{equation}
Here, the decomposition coefficients $B_k$---or ``bucket values'', in ghost-imaging language \cite{bromberg2009ghost,katz2009compressive}---are defined via the inner product (dot product)
\begin{equation}\label{eq:VanillaGhostProjectionBucketCoefficients}
B_k :=\frac{1}{\Upsilon}\iint I_k(\mathbf{r})S(\mathbf{r}) \,d\mathbf{r} =: I_k \cdot S
\end{equation}
and
\begin{equation}
 \overline{S} := \frac{1}{\Upsilon} \iint S(\mathbf{r})\, d\mathbf{r}. 
\end{equation}
The right side of Eq.~(\ref{eq:VanillaGhostProjection}) shows how one may form a suitable linear combination of the previously mentioned speckle maps, to give a smoothed version of $S(\mathbf{r})$ that is convolved with the point-spread function (smeared Dirac delta) associated with the smoothed completeness relation.  The speckle patterns $I_k(\mathbf{r})$ have the following properties: the spatial average $\overline{I_k}$ is approximately constant with respect to $k$ (to indicate that it is independent of $k$, we shall write this number as $\bar{I}$), and the ensemble average $\langle I(\mathbf{r})\rangle$ is approximately uniform with respect to $\mathbf{r}$, equal to $\bar{I}$ everywhere. Using these assumptions and expanding Eq.~(\ref{eq:VanillaGhostProjection}), we obtain
%
\begin{equation}\label{eq:VanillaGhostProjection2}
[S(\mathbf{r})\otimes PSF(\mathbf{r}) ]+ \langle B_k \rangle \bar{I}= \left\langle B_k I_k(\mathbf{r})\right\rangle.
\end{equation}
This result is illustrated in Fig.~\ref{fig:VanillaGhostProjection}. It may be spoken of as ``building signals out of noise'' \cite{paganin2019writing}, in the sense that it expresses the smoothed target pattern, $S(\mathbf{r})\otimes PSF(\mathbf{r})$, plus the offset (pedestal), $ \langle{B_k}\rangle \bar{I}$, via a linear combination of speckle (noise) maps, $\langle B_k I_k(\mathbf{r})\rangle$.


\section{Diffraction catastrophes as a random basis for ghost projection}\label{sec:DiffCatastrophesAsRandomBasis}

In this paper, we propose an alternative basis for ghost projection, replacing the ensemble of speckle maps with an ensemble of focal-field diffraction catastrophes \cite{kravtsov2012caustics}.\footnote{If the aberrations associated with the distorted focal field are sufficiently strong, the resulting focal-plane caustics constitute the ``diffraction catastrophes'' \cite{kravtsov2012caustics, Nye1999} referred to in this sentence from the main text.  If the aberrations are sufficiently weak---while also being large enough to split a focal spot into a relatively large number of local intensity maxima---the peaked focal-field caustic intensity structure will be significantly softened.  In this latter case, one will instead have distorted arrays of diffraction spots such as those illustrated on p.~541 of \citet{born2013principles}. The methods of our paper may also be applied to such a case, namely where the aberrated focal-field patterns are highly structured, but the aberrations are not sufficiently strong for focal-plane caustics to be evident. A similar statement holds for a complementary limiting case, where the geometrical-optics limit implies that caustic-laden focal-field intensities may be described using ray optics.} This corresponds to modifying the scenario in Fig.~\ref{fig:VanillaGhostProjection} to that shown in Fig.~\ref{fig:Generating diffraction catastrophes}(b). The key changes are that the correlation length $\ell$ is now a function of transverse position $\mathbf{r}$ and the pedestal, $\mathcal{P}(\mathbf{r})$, is no longer spatially uniform. Equation~(\ref{eq:VanillaGhostProjection2}) thereby generalizes to
\begin{equation}\label{eq:CatastropheGhostProjection}
\left[\iint S(\mathbf{r}') PSF(\mathbf{r},\mathbf{r}')\,d\mathbf{r}'\right]+ \mathcal{P}(\mathbf{r})= \langle B_k I_k(\mathbf{r})\rangle,
\end{equation}
whereby the space-invariant point-spread function $PSF(\mathbf{r})$ has been replaced with a space-variant smearing function (scaled autocovariance) $PSF(\mathbf{r},\mathbf{r}')$ that converts the convolution into a linear integral transform, and $\{I_k(\mathbf{r})\}$ now denotes an ensemble of diffraction catastrophes.  In practice, the position dependence of the smearing function $PSF(\mathbf{r},\mathbf{r}')$ will often be fairly weak, if one restricts the field of view for the desired ghost projection to be in a sufficiently small vicinity of the optical axis.

From this point we will utilize a vector-space description of pixellated images to lay the foundation for understanding how ghost projection functions in practice, and how it can be optimized. Notions of superposition and weighting of patterns naturally lead to treating the illuminating patterns as elements of a vector space \cite{BarrettMyersBook}, a formalism which is amenable to simulation and creating weight-generating algorithms. In ghost projection, the set of illuminating patterns can be considered as a fixed set of function-space vectors; the weight-generating aspect of ghost projection is analogous to finding a linear combination of these vectors that approximates some desired vector (image), with the constraint that the coefficients be non-negative because negative exposure times are non-physical. A consequence of this non-negativity constraint is that the image can only be synthesized up to the previously mentioned offset that we refer to as the ``pedestal'' \cite{paganin2019writing,ceddia2022ghost}. 

While the illuminating patterns are elements of an infinite-dimensional function space \cite{BarrettMyersBook}, by reducing each pattern to a raster image, the patterns can be approximated by vectors whose entries represent the values (intensities) of each pixel \cite{ceddia2022ghostII}. This approximation is valuable for computational considerations. These raster-image vectors live in a smaller function space, which has a number of dimensions equal to the number of pixels per image.

Let the desired ghost-projection image be $S$, and the set of weights for the illuminating patterns be $\{w_k\}_{k \in \{1,\cdots,\mathcal{N}\}}$, where each $w_k \geq 0$ and each $I_k$ has only non-negative entries. Note that for ease of notation, we have suppressed the functional dependence on $\mathbf{r}$. The goal of the computational side of ghost projection is to find some optimum set of weights, $w_k$, such that in combination they provide an accurate synthesis of $S$ \cite{paganin2019writing, ceddia2022ghost, ceddia2022ghostII}, i.e.,
\begin{equation}
    \sum_{k=1}^{\mathcal{N}} w_k I_k \approx S + \mathcal{P}.
\end{equation}

The scheme for synthesizing a ghost projection (up to an additive pedestal and the previously-indicated position-dependent smearing) will often be rather inefficient.  It is at this point that one can make constructive use of the fact that one need only (in fact, must only) choose a vanishingly small fraction of the set of all possible focal-field diffraction catastrophes, in performing the ghost-projection synthesis given by the right side of Eq.~(\ref{eq:CatastropheGhostProjection}). The methods to choose diffraction catastrophes and determine appropriate weightings that have been proposed in the literature to date, ordered in terms of increasing computational complexity, are as follows:
\begin{enumerate}[(i)]

    \item Our first ghost-projection method is the entirely-analytic albeit highly inefficient approach of Ref.~\cite{paganin2019writing} that is summarized in Eqs.~(\ref{eq:VanillaGhostProjectionBucketCoefficients}) and (\ref{eq:VanillaGhostProjection2}) above. Weights are generated via the dot product, $B^{\prime}_k$, as follows:
    \begin{equation}\label{eqn:dotProduct}
        w_k = \left\{ 
            \begin{array}{ c l }
            B^{\prime}_k & \quad \textrm{if } B^{\prime}_k > 0\\
            0   & \quad \textrm{otherwise}
            \end{array}
          \right. \quad \textrm{where} \quad B^{\prime}_k = I_k \cdot S'.
    \end{equation}
    Here $S'=S-\overline{S}$ is the mean-corrected form of $S$.
    
    \item The efficiency of this ``dot product'' ghost-projection scheme can be improved by increasing the selection threshold from zero to $\epsilon > 0$ in Eq.~(\ref{eqn:dotProduct}), as follows:
    \begin{equation}\label{eqn:dotProduct2}
        w_k = \left\{ 
            \begin{array}{ c l }
            B^{\prime}_k & \quad \textrm{if } B^{\prime}_k \geq \epsilon\\
            0   & \quad \textrm{otherwise.}
            \end{array}
          \right.
    \end{equation}
    This modification enforces sparsity in the non-zero weightings, keeping only those members of the basis that have a particularly high cross-correlation with the target image. However, a larger $\epsilon$ can lead to the filtration of a prohibitively large fraction of the candidate random-basis elements \cite{paganin2019writing}.

    \item A significantly more efficient method of synthesizing the desired image, $S$, is to use a suitable iterative numerical method such as the non-negative least squares (NNLS) algorithm \cite{ceddia2022ghost, ceddia2022ghostII}. We recall the previously mentioned fact that the weighting coefficients $w_k$ cannot be negative, since they are proportional to the exposure times for each element of the random basis. NNLS can be employed to determine the set $\{w_k\ge 0\}$ that minimizes the cost function 
    \begin{equation}\label{eq:NNLSghostProjection1}
        \mathcal{C}=\left\lVert \sum_{k=0}^{\mathcal{N}} w_k I_k - S \right\rVert_2,
    \end{equation}
    where $\lVert \cdot \rVert_2$ denotes the Euclidean norm. This can be solved iteratively using the method in Chap.~23 of the book by Lawson and Hanson \cite{LawsonHanson1995}. In this work we have utilized the python function {\it scipy.optimize.nnls} to determine the weights.
    
\end{enumerate}

The optimal solution to Eq.~(\ref{eq:NNLSghostProjection1}) will choose a minimal set of basis patterns to avoid adding a pedestal to the result that has a square-error cost. Depending on the size of the basis set, the constructed image may not have a high quality in terms of reproducing the target image details. Should the application allow it, we can improve image fidelity by allowing a flat pedestal to be added. An effective method to achieve that is to determine the non-negative weights $w_k$ that minimize
\begin{equation}\label{eq:NNLSghostProjection2-MC}
    \mathcal{C}=\left\lVert \sum_{k=0}^{\mathcal{N}} w_k I'_k - S' \right\rVert_2,
\end{equation}
where $I_k'(\mathbf{r}):=I_k(\mathbf{r})-\overline{I_k}$ and $S'(\mathbf{r}):=S(\mathbf{r})-\overline{S}$ are mean-corrected forms of $I_k$ and $S$. This is ideal when the ensemble-averaged pattern $\langle I(\mathbf{r})\rangle$ (henceforth abbreviated as the AIP for \textit{average illuminating pattern}) is relatively flat. Further modifications of this optimization are required when the AIP is not flat (as is the case with diffraction catastrophes). This will be developed in Sec.~\ref{sec:results_modifying_target}.

While approach (ii) enforces sparsity in an explicit manner, approach (iii) implicitly imposes sparsity, since NNLS typically assigns significant weights $w_k$ to only a small fraction of the random basis \cite{Gorban2016} (although this does depend on the target image).
As a result, if we discard weights below a specified threshold, the number of patterns projected can be greatly reduced with minimal detriment to reconstruction quality. The larger the number of members in the random basis from which one can choose, the more efficient the ghost-projection process can become.\footnote{We have formalized several schemes to select the sparse set of non-zero pattern weightings to produce a desired image, given an initial large set of basis patterns. These schemes will be described later in the paper.} 

\section{Simulation of focal-field diffraction catastrophes}\label{sec:MethodsMainHeading}

Here we outline our means for simulating diffraction catastrophes. First, random phase modulations are constructed via Zernike polynomials, as explained in Sec.~\ref{sec:ZP}. The random phase maps are then focused to produce diffraction catastrophes, as described in Sec.~\ref{sec:focusing}. In Sec.~\ref{sec:noise} we outline how realistic photon shot-noise is added to the simulations. Here we also define a suitable signal-to-noise ratio (SNR) metric.

\subsection{Truncated Zernike basis for phase modulations}
\label{sec:ZP}

The beam in Fig.~\ref{fig:Generating diffraction catastrophes}(a) is taken to be transversely confined by a circular aperture. Zernike polynomials \cite{born2013principles,Noll1976,WangSilva1980,lakshminarayanan2011zernike,niu2022zernike} form a complete orthogonal set over such a region. Additionally, the low-order Zernike polynomials have low spatial frequency. They therefore provide a convenient basis for describing the slowly-varying phase modulations that can be used to generate diffraction catastrophes for the purposes of ghost projection. Working in plane polar coordinates $(\rho,\theta)$, denote the first 21 Zernike polynomials by
\begin{equation}
\mathcal{Z} := \{Z_1(\rho,\theta),Z_2(\rho,\theta),\cdots,Z_{21}(\rho,\theta)\}. 
\end{equation}
We work on the unit disk $\rho \le 1$, and use the standard convention in \citet{Noll1976} and \citet{WangSilva1980} for indexing the polynomials. 

Let $\{\psi_k\}_{k \in \{1,\cdots,\mathcal{N}\}}$ be the set of phase-modulated wavefunctions before focusing (defined on the cross section of the beam), where $\mathcal{N}$ is the number of illuminating patterns used for a given reconstruction. Each (complex-valued) wavefunction $\psi_k$ has a (real-valued) phase function $\phi_k$, defined as a linear combination of Zernike polynomials from $\mathcal{Z}$. Let the coefficient of $Z_j$ be $c^k_j \in \mathbb{R}$,
i.e.
\begin{equation}
\phi_k(\rho,\theta) = \sum_{j=2}^{21} c^k_j Z_j(\rho,\theta).    
\end{equation}
Note that the $j=1$ term is omitted in the summation, since it corresponds to a constant additive phase that has no physical meaning in the present context.

To bound the phase modulations, each $c^k_j$ is randomly chosen from the interval $[-\tau,\tau]$ for some positive real value $\tau$. In order for the focal-field intensity to contain a caustic network, the value for $\tau$ needs to be sufficiently large.  We set the amplitude of the modulated wavefunction to be uniformly 1 within the unit circle and zero elsewhere, thereby simulating a circular aperture that is uniformly illuminated by normally-incident strictly monochromatic plane waves, so
\begin{equation}
\psi_k(\rho,\theta) =
    \begin{cases}
        e^{i\phi_k(\rho, \theta)} \quad & \text{if } \rho \leq 1 \\
        0 \quad & \text{otherwise}.
    \end{cases}
\end{equation}

For most of the results in this paper (from Sec.~\ref{sec:results_modifying_target} onwards), a set of $\mathcal{N} = 7500$ randomly-generated illuminating patterns was used to construct $100 \times 100$ pixel target images. 
Naturally, $\mathcal{N} =7500$ patterns cannot span the $10\,000$-dimensional space of $100\times 100$ pixel images (especially when restricted to only using non-negative coefficients). It is worth noting that this means the set of patterns does not strictly form a basis in the mathematical sense (a linearly independent spanning set). Nonetheless, with the target image modifications explored in Sec.~\ref{sec:results_modifying_target}, a set of 7500 patterns is ample to produce successful reconstructions, emphasizing the effectiveness of this technique. In practice, assuming there are no constraints on computation, larger values of $\mathcal{N}$ lead to more successful reconstructions. If $\mathcal{N}$ is far greater than the number of pixels, most vectors will be able to be constructed by numerous different non-negative linear combinations of the illuminating patterns, although these linear combinations may differ in terms of the magnitude of the pedestal they produce. Consequently, as an area of further exploration, it makes sense to talk about choosing the linear combination which maximizes the signal-to-pedestal ratio (which leads to an improved signal-to-noise ratio in the presence of photon shot noise). In this paper, no mechanism for this has been explored. 

For $k \in \{1,\cdots,\mathcal{N}\}$ and $j \in \{2,\cdots,21\}$, each of the $c^k_j$ was uniformly randomly selected from $[-\tau,\tau]$. The size of $\tau$ was found to be approximately proportional to the average size of the corresponding focal-field illuminating pattern, which is natural on account of the fact that the mean aperture-plane phase gradient is linearly proportional to $\tau$. For this paper, we choose $\tau = 9$, as for $100 \times 100$ pixel images, this makes the patterns a suitable size for the target images to be synthesized.

\subsection{Focusing the phase modulations via the Fourier transform}
\label{sec:focusing}

A coherent two-dimensional radiation distribution, when propagated infinitely far, becomes the Fourier transform of the initial distribution. A focusing lens enables this to occur at a finite propagation distance, and the Fourier transform appears at the focal plane \cite{born2013principles}. In particular, once the modulated wavefunction, $\psi_k$, is focused via a lens, the intensity distribution, $I_k$, in the focal plane is given by 
\begin{equation}\label{eqn:intensity}
I_k = \left| \mathcal{F} \left[\psi_k \right]\right|^2 = \left| \mathcal{F} \left[ A(\rho) e^{i\phi_k(\rho,\theta)} \right]\right|^2.
\end{equation}
Here, $\mathcal{F}$ denotes Fourier transformation with respect to transverse Cartesian coordinates $\rho\cos\theta$ and $\rho\sin\theta$, and $A(\rho)$ is an aperture function in scaled (dimensionless) spatial coordinates, defined to be unity for $\rho \leq 1$ and zero elsewhere. The resulting $I_k$ are diffraction catastrophes \cite{PostonStewartTheoryBook, Nye1999, kravtsov2012caustics}. Here, $\{I_k\}_{k \in \{1,\cdots,\mathcal{N}\}}$ serves as the caustic-laden ``catastrophe basis'' set of illuminating patterns to be used for ghost projection. The process described above is depicted in Fig.~\ref{fig:Generating diffraction catastrophes}(a) and some example patterns are shown in Fig.~\ref{fig:Generating diffraction catastrophes}(b). For simulation purposes, it is helpful to rasterize the wavefunctions, treating $A(\rho)\exp[i\phi_k(\rho, \theta)]$ as a vector with a complex-valued entry for each pixel. This allows the Fourier transform step to be performed computationally as a 2D fast Fourier transform (FFT) \cite{Press2007}.  To avoid wrapping of the FFT at the image boundaries, the phase function must be zero-padded \cite{Press2007}, i.e., the phase function is given a sufficiently broad border of value zero.

\subsection{Simulating Poisson noise in the focal fields}
\label{sec:noise}

The number of photons (or electrons, neutrons, etc.) incident on a pixel of the projection plane obeys a Poisson probability distribution \cite{MandelWolf, LoudonBook}. For simplicity, assume that the projecting beam is composed of photons. For the $j$th pixel of the ghost projection, let $\Lambda_j := c \sum_k w_k [I_k]_j$ represent the expected number of photons incident on that pixel for some scaling factor $c$. When measured, the number of photons incident at each pixel, $j$, will be Poisson distributed, with mean $\Lambda_j$ and standard deviation $\sqrt{\Lambda_j}$. To quantify noise levels, a pixel-wise signal-to-noise ratio (SNR$_j$) is introduced, given by
\begin{equation}
\text{SNR}_j := \frac{|\Lambda_j-P|}{\sqrt{\Lambda_j}}
\end{equation}
where $P$ is the average number of photons per pixel of the pedestal. Note that $|\Lambda_j - P|$ is the deviation of the projection from the pedestal at pixel $j$, representing useful signal. Defining the global SNR for an $N \times N$ pixel image as the root-mean square (RMS) of the pixel-wise SNR$_j$, we have
\begin{equation}\label{eq:global SNR}
    \text{SNR} := \sqrt{\sum_{j=0}^{N^2} \frac{\text{SNR}_j^2}{N^2}} = \frac{1}{N} \left\lVert \frac{\Lambda - P}{\sqrt{\Lambda}} \right\rVert_2.
\end{equation}
Note that this definition yields lower SNR values if the target image is 0 (i.e.~$\Lambda = P$) for many of the pixels, even if the noise level and maximum signal is kept constant. 

\section{Results and Analysis}\label{sec:Results and analysis}

In this section we first compare ghost projection performance using both (i) the analytical dot product method, and (ii) the iterative non-negative least-squares method, to determine pattern weights  in Sec.~\ref{sec:results_synthesis}. Unlike a random speckle-pattern basis that produces a flat (i.e. uniform) pedestal in target-image synthesis, we observe that a diffraction-catastrophe basis produces a non-uniform pedestal.  This observation is presented in Sec.~\ref{sect:results_nonuniform_pedestal}. We then present a method to modify the target image to compensate for this non-uniform pedestal in Sec.~\ref{sec:results_modifying_target}. Finally, we explore the effects of Poisson noise on target image quality in Sec.~\ref{sect:results_noise}.

\subsection{Synthesis of unmodified target images}
\label{sec:results_synthesis}

\begin{figure}[!htb]
    \centering
    \includegraphics[width=\linewidth]{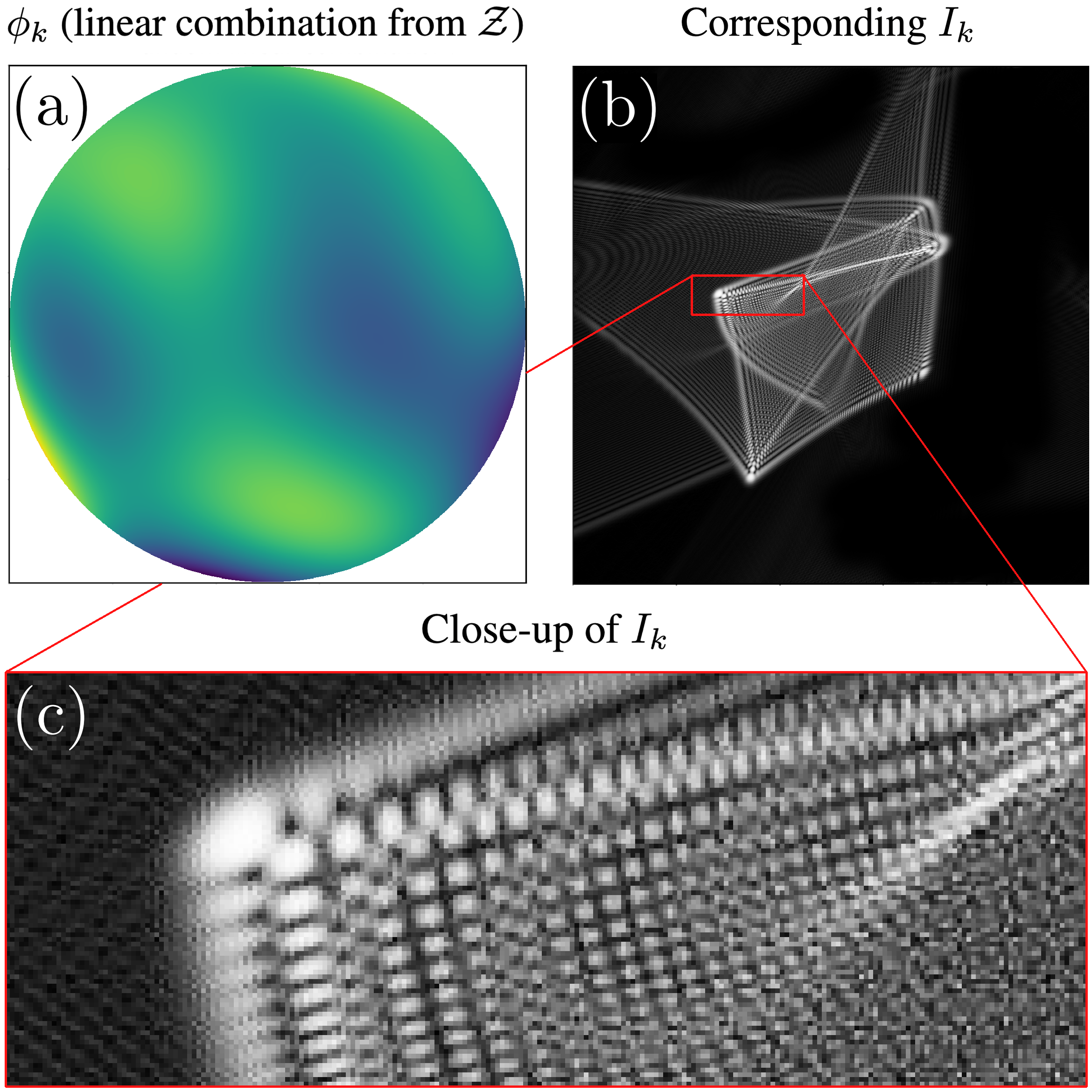}
    \caption{(a) A phase function $\phi_k$ generated by a random linear combination of the 2nd to 21st Zernike polynomials, (b) the corresponding diffraction-catastrophe illuminating pattern $I_k$, (c) a zoomed region of $I_k$.
    }
    \label{fig:Example pattern}
\end{figure}

Figure~\ref{fig:Example pattern}(a) shows an example of a phase function, $\phi_k$, that is generated randomly using Zernike polynomials from $\mathcal{Z}$, with the corresponding diffraction catastrophe, $I_k$, after focusing shown in Figs.~\ref{fig:Example pattern}(b-c).  The spatial scale is in arbitrary units, for all panels. The caustic diffraction catastrophes \cite{kravtsov2012caustics} are evident as the bright lines in the intensity images. While these lines would correspond to nonphysical lines of infinite intensity in the geometric-optics limit of zero wavelength, the finite wavelength softens them to peaked intensities decorated with fine grid-like structures associated with a lattice of alternating phase vortices and antivortices \cite{Nye1999}. 

\begin{figure}[!htb]
    \centering
    \includegraphics[width=\linewidth]{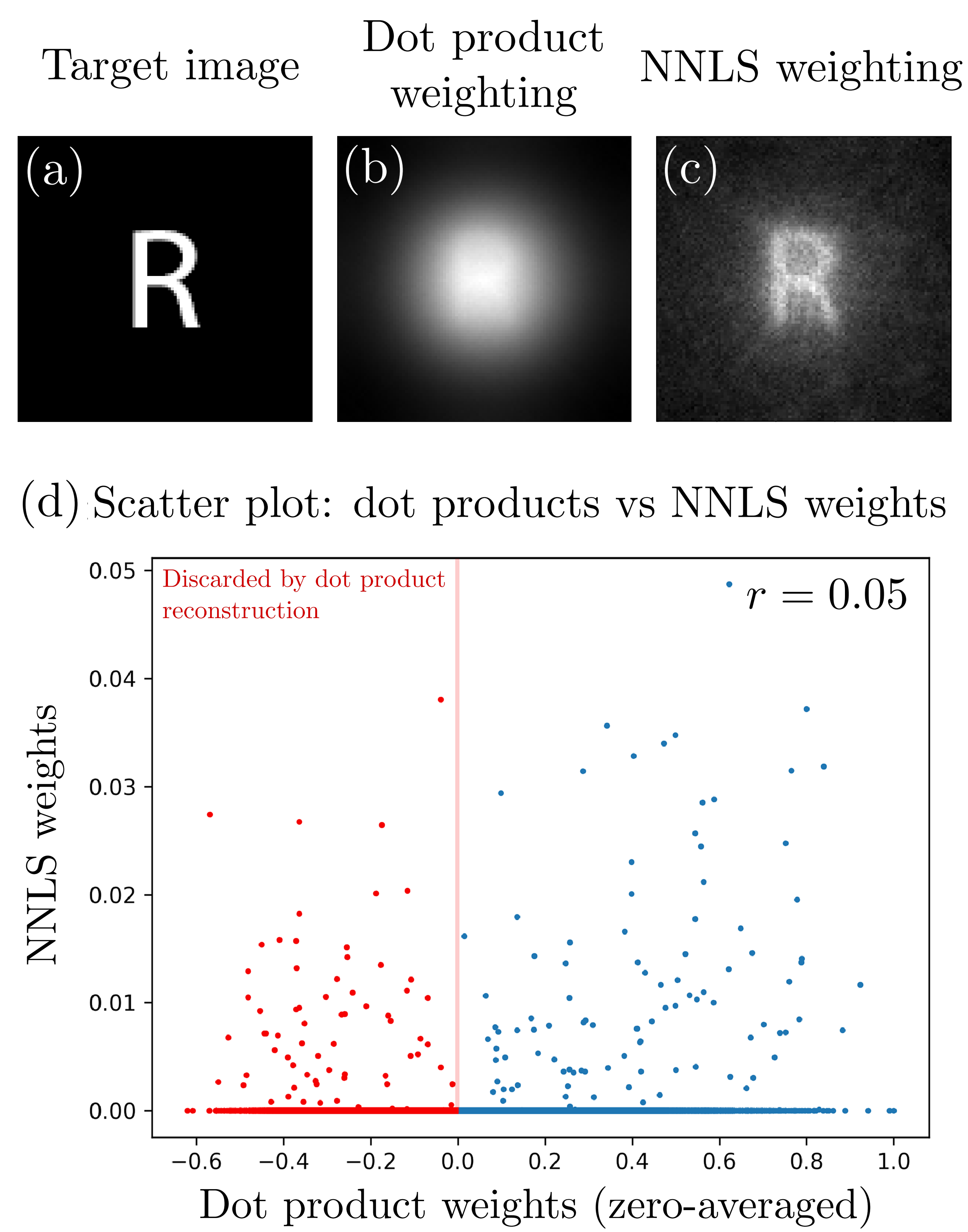}
    \caption{ (a) A $100 \times 100$ pixel image of the letter ``R'' used as a target image, (b) a simulated ghost-projection reconstruction of the target image using 40\,000 patterns, weighting the catastrophes by their dot product with the target image, (c) a reconstruction using the non-negative least squares algorithm to generate weights, using 40\,000 illuminating patterns. The reconstructions have been normalized, so that the minimum value is rendered black and the maximum value is rendered white. (d) There is low correlation between the two methods for generating the weights, $w_k$, with the Pearson correlation coefficient $r$ being 0.05. }
    \label{fig:Normal reconstructions}
\end{figure}

Using a set of $\mathcal{N} = 40\,000$ randomly generated aberrated focal-field diffraction-catastrophe patterns as a random basis, the simulated ghost projections in Fig.~\ref{fig:Normal reconstructions} are produced.  The target image, here, is the binary capital letter ``R'' in Fig.~\ref{fig:Normal reconstructions}(a).  Two different ghost-projection algorithms are explored below, namely dot product ghost projection \cite{paganin2019writing} and NNLS ghost projection \cite{ceddia2022ghost, ceddia2022ghostII, ceddia2023universal}, as described in the following paragraphs (see, also, Sec.~\ref{sec:DiffCatastrophesAsRandomBasis}). The dot product reconstruction in Fig.~\ref{fig:Normal reconstructions}(b) has comparatively far lower contrast, with the white section of the target image sitting only approximately 5\% above the local pedestal, whereas in the NNLS construction of Fig.~\ref{fig:Normal reconstructions}(c), the white section sits approximately 34\% above the local pedestal. This greater contrast is particularly valuable when noise is present (as explored further in Sec.~\ref{sec:noise}).

The ``dot product'' ghost-projection construction, which follows Ref.~\cite{paganin2019writing} by using weights defined in Eq.~(\ref{eqn:dotProduct}), discards illuminating patterns that have a below-average dot product with the target image. If one thinks of the random illumination patterns as randomly directed vectors in a function space, this method for ghost projection has the simple geometric interpretation that any function-space vector may be synthesized by taking a sheaf of random vectors and retaining only those having a positive projection along the direction of the required vector; the sum of these retained vectors is then approximately parallel to the desired vector (target ghost-projection image).\footnote{This geometric construction is illustrated in Fig.~11 of Ref.~\cite{paganin2019writing}.} Illuminating patterns with lower dot products with the desired image increase the pedestal while having a reduced contribution towards improving the ghost-projection image. Hence the weights of the included illuminating patterns are the dot product between the image and illuminating pattern vector, minus the average dot product. While conceptually clear, the very low contrast of the ``dot product'' ghost-projection method (see Fig.~\ref{fig:Normal reconstructions}(b)) argues against its utility in practice.

Employing the NNLS-based method for ghost projection \cite{ceddia2022ghost, ceddia2022ghostII, ceddia2023universal}, as specified by Eq.~(\ref{eq:NNLSghostProjection2-MC}), we obtain the simulated catastrophe-basis ghost projection in Fig.~\ref{fig:Normal reconstructions}(c).  A significant improvement in contrast is evident.  Note that the NNLS-based method employs basis members having both above-average and below-average dot product with the target image.  Conceptually, this corresponds to background-subtracted basis members that are either positively or negatively correlated with the target image, with the inclusion of the latter set amounting to a form of ``eraser'' that complements the structured delocalized ``pencils'' of the former set.

Further comparison of the two ghost-projection approaches may be obtained by plotting each weight, $w_k$, obtained using the NNLS approach, versus the corresponding weight obtained via the dot product approach. The resulting scatter plot in Fig.~\ref{fig:Normal reconstructions}(d) reveals a notable disparity between the two reconstruction algorithms, with no significant correlation between the weights $w_k$ generated by each procedure. While Pearson's correlation coefficient ($r = 0.05$, stated in Fig.~\ref{fig:Normal reconstructions}(d)) is positive, the support for a linear relationship between the dot product and the NNLS weights is weak, so the dot product cannot easily be used to filter illuminating patterns before the NNLS algorithm is applied. It is worth noting that the dot-product weights are not symmetrically distributed around their mean and are positively skewed. Additionally, the NNLS algorithm only assigns 158 illuminating patterns a weight that is over 1\% of the maximum weight (only 0.395\% of the 40\,000 illuminating patterns), with 62 of these having a dot product below the mean correlation (red) and 96 having a dot product above the mean correlation (blue). This small selection of contributing illuminating patterns is the reason that the NNLS image is noticeably more grainy than the dot product reconstruction, without a large quantity of similarly-weighted illuminating patterns to smooth the image.  

\subsection{Non-uniform pedestal}
\label{sect:results_nonuniform_pedestal}

\begin{figure}[!htb]
    \centering
    \includegraphics[width=\linewidth]{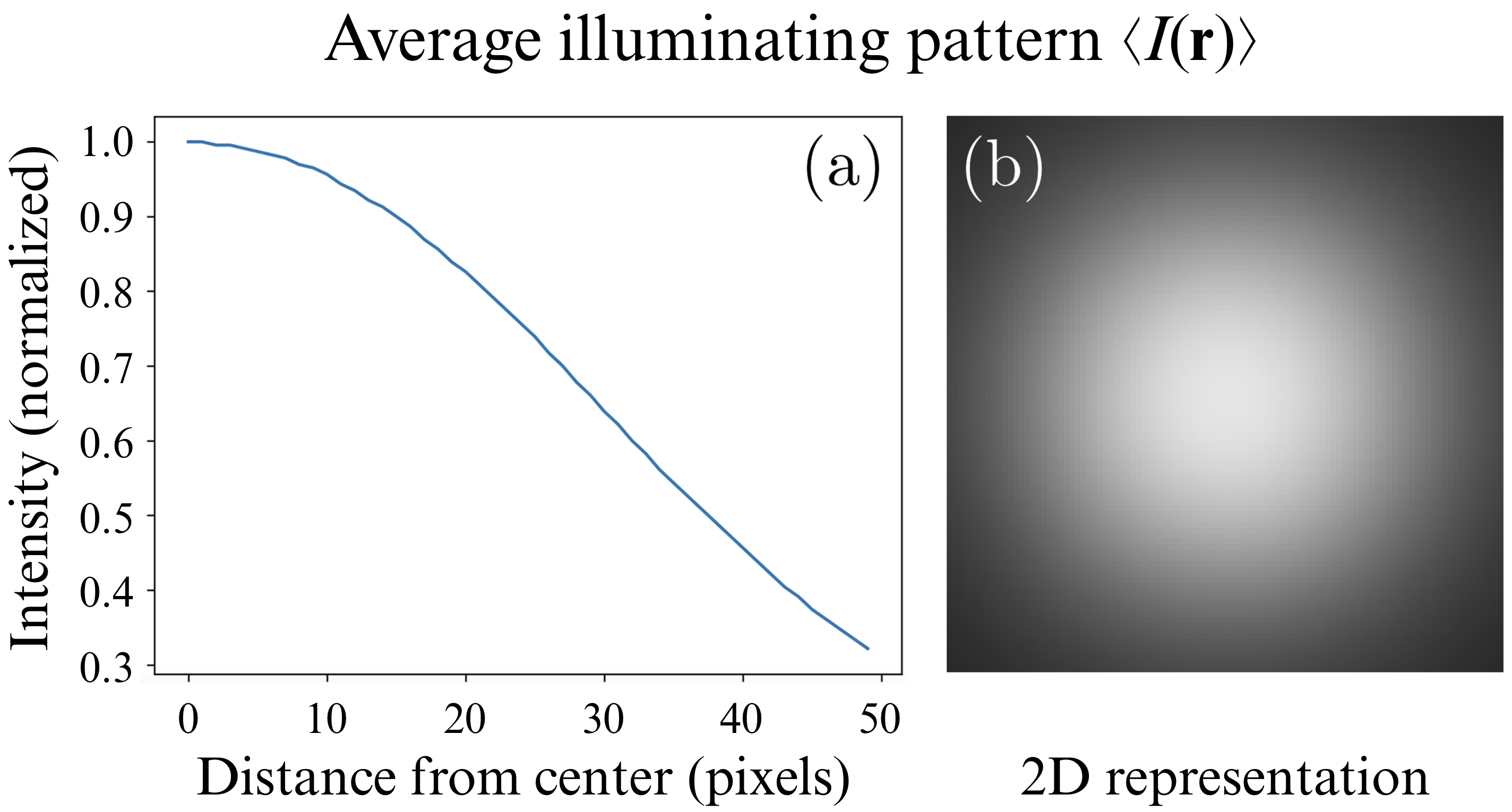}
    \caption{The average diffraction catastrophe, plotted as (a) a 1D profile versus radial distance and (b) a 2D grayscale image. The 2D image above has not been normalized. As may be expected from the central limit theorem, the average illuminating pattern is well approximated by a Gaussian distribution. }
    \label{fig:Pedestal}
\end{figure}

Another significant aspect of the images formed in Fig.~\ref{fig:Normal reconstructions} is the additional background signal, or pedestal; in particular, that it is not spatially uniform. For the case of the dot product reconstruction in Fig.~\ref{fig:Normal reconstructions}(b), the pedestal is approximately the average illuminating pattern (see Fig.~\ref{fig:Pedestal}), while the NNLS construction in Fig.~\ref{fig:Normal reconstructions}(c) has a flatter pedestal of similar profile (as there are far fewer contributing illuminating patterns, the resulting pedestal when using NNLS is more dependent on the target image). 

Reconstruction is most effective when the target image is only a small perturbation away from the average illuminating pattern (AIP) in Fig.~\ref{fig:Pedestal}. If the mean-corrected target image exhibits only small deviations from zero (i.e.~it is relatively flat), adding the AIP to the target image produces a \textit{modified target image} which is only a small perturbation away from the AIP, as desired. Observing Eq.~(\ref{eq:NNLSghostProjection2-MC}), adding the AIP to the target image achieves a similar effect to replacing the patterns $I_k(\mathbf{r})$ in Eq.~(\ref{eq:NNLSghostProjection1}) with $I_k(\mathbf{r}) - \langle I(\mathbf{r})\rangle$. This is a more sophisticated version of the mean correction in Eq.~(\ref{eq:NNLSghostProjection2-MC}), with the patterns shifted by the AIP rather than a constant value. 

This mean correction approach is suboptimal when noise is present (since it produces large pedestals) or when the target image is not relatively flat. In the next section, we will explore a more general tunable technique to account for these scenarios. 



\subsection{Modified target images}
\label{sec:results_modifying_target}



As previous mentioned, the target image $S(\mathbf{r})$ is reconstructed most successfully when its overall spatial profile is similar to $\langle I(\mathbf{r})\rangle$, the AIP. Intuitively, this is because many more patterns will already share large-scale features with the target image, allowing the NNLS to focus on small deviations and prioritize reconstruction of finer details.

\begin{figure}[!htb]
    \centering
    \includegraphics[width=\linewidth]{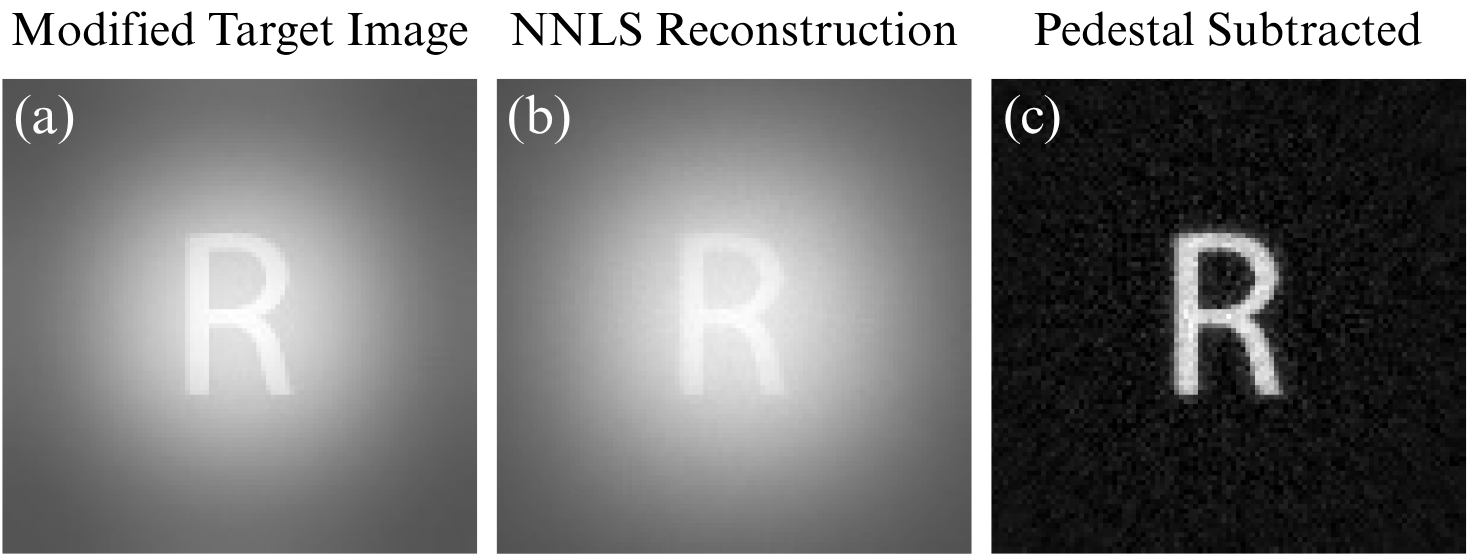}
    \caption{ (a) Modified target image of the capital letter ``R'' using $\beta = 0.1$ in Eq.~(\ref{eq:beta}), (b) NNLS reconstruction of this modified target image using 7500 illuminating patterns, (c) reconstruction after the pedestal, namely $(1-\beta)\langle I(\mathbf{r})\rangle$, has been subtracted. }
    \label{fig:Reconstruction with pedestal added prior}
\end{figure}

To exploit this fact, we propose a simple method for modifying the target image to be more similar to the AIP. Specifically, instead of projecting the target image $S(\mathbf{r})$, we project
\begin{equation}\label{eq:beta}
S_\beta(\mathbf{r}):=\beta S(\mathbf{r}) + (1-\beta)\langle I(\mathbf{r})\rangle,
\end{equation}
where $\beta \in (0,1]$ is a tunable parameter. Note that $S_1(\mathbf{r})=S(\mathbf{r})$. An example of this has been presented in Fig.~\ref{fig:Reconstruction with pedestal added prior} for $\beta = 0.1$. Naturally, when $\beta\neq1$, it is necessary to ``subtract'' $(1-\beta)\langle I(\mathbf{r})\rangle$, referred to as the ``pedestal'', afterwards to arrive at the desired final image. Methods for this subtraction (or ``pedestal flattening'') are explored in Sec.~\ref{sec:PracticalConsiderationsAndFuturework}.


\begin{figure}
    \centering
    \includegraphics[width=\linewidth]{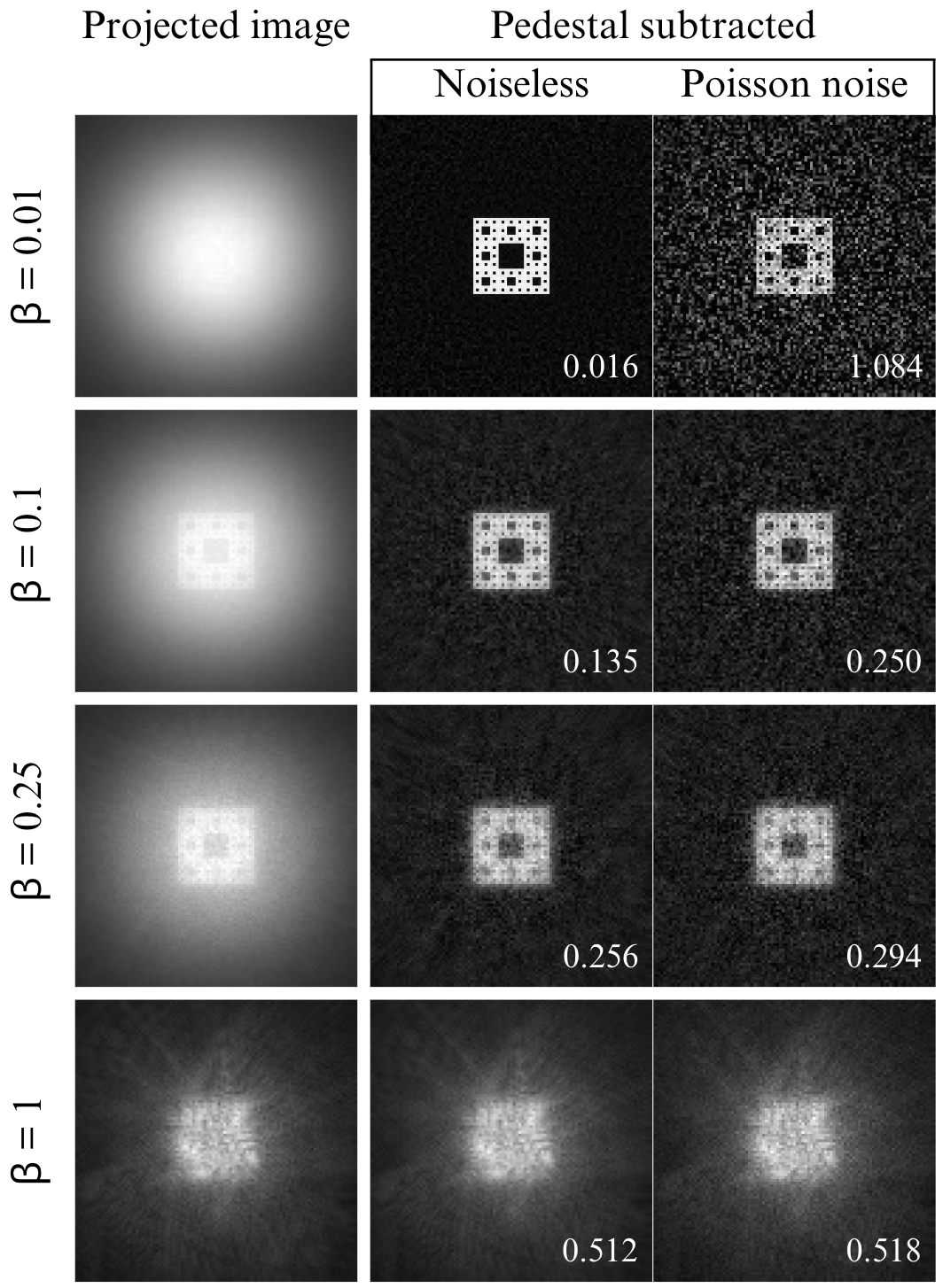}
    \caption{NNLS reconstructions of a Sierpi\'{n}ski carpet using different $\beta$ values (see Eq.~(\ref{eq:beta})). Dimensionless residuals are included in the bottom right corner of each image. To simulate Poisson noise (photon shot noise), the projection's exposure is such that, on average, a maximum-exposure pixel of the target image receives 1000 photons more than a minimum-exposure pixel at the same location. The total number of photons hitting a pixel at position $\mathbf{r}$ is $N(\mathbf{r}) = \frac{1000}{\beta}(1-\beta)\langle I(\mathbf{r})\rangle + 1000S(\mathbf{r})$. Since the standard deviation of the Poisson noise is $\sqrt{N(\mathbf{r})}$, the SNR is lower for smaller $\beta$. There is thus a compromise between SNR and reconstruction quality, as illustrated in the third column.  }
    \label{fig:different betas}
\end{figure}

\begin{figure}[h]
    \centering
    \includegraphics[width = \linewidth]{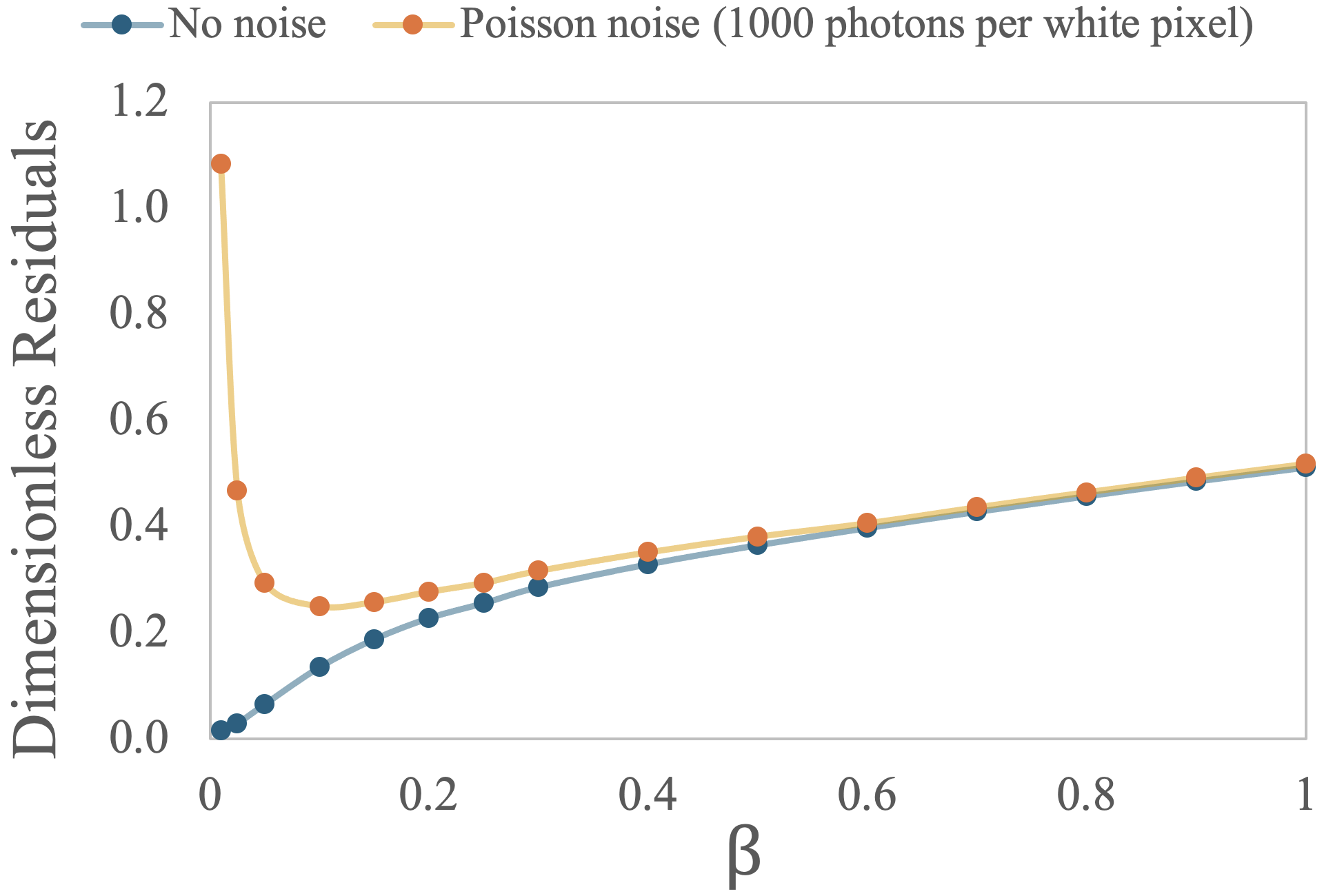}
    \caption{Dimensionless residuals of reconstructions of a Sierpi\'{n}ski carpet as a function of $\beta$. In the absence of noise (blue), optimum values are typically $\beta\sim10^{-2}$. However, Poisson noise (photon shot noise) is more impactful for small $\beta$. To simulate the Poisson noise (orange), the projection exposure is set such that a maximum-exposure pixel (``white pixel'') of the original target image receives an average of 1000 photons more than a minimum-exposure pixel at the same location. This leads to an optimum of $\beta \approx 0.1$.}
    \label{fig:Residuals vs beta}
\end{figure}

The extreme values for $\beta$ do not lead to optimal results: $\beta = 0$ does not have any contribution from the target image, while $\beta = 1$ corresponds to the unmodified target image, which is suboptimal in most scenarios (as illustrated in Figs.~\ref{fig:different betas} and \ref{fig:Residuals vs beta}). To quantify and compare reconstruction quality, we define ``dimensionless residuals'' as the norm squared of the reconstruction (after pedestal subtraction) minus the target image divided by the norm squared of the target image.

It is possible to optimize the value of $\beta$ by allowing the NNLS algorithm to adjust it as one of its parameters (this typically results in a low-contrast image reconstruction with $\beta \sim 10^{-2}$; the target image is reduced to a slight perturbation about the AIP). To achieve this, the negative of the AIP can be included with the illuminating patterns when the NNLS is conducted, with $\beta$ simultaneously optimized along with the set of exposure times for each pattern (equivalent to setting $\beta = w_{\mathcal{N}+1}$ and $-\langle I(\mathbf{r}) \rangle = I_{\mathcal{N}+1}$ in Eq.~(\ref{eq:NNLSghostProjection2-MC})). Although this leads to a mathematically optimal choice of $\beta$ (according to the Euclidean norm), small values of $\beta$ come at the cost of a lower signal-to-pedestal ratio, making the reconstruction more vulnerable to noise in an experimental realization of ghost projection. As a result, choosing $\beta$ is a matter of balancing artifacts with noise. Figure~\ref{fig:Residuals vs beta} demonstrates an example of this compromise in the presence of Poisson noise, where low values of $\beta$ lead to poor dimensionless residuals due to having low signal-to-noise ratio, while large values of $\beta$ lead to unwanted reconstruction artifacts, resulting in $\beta \approx 0.1$ having a minimal residual. The optimum $\beta$ for any given application will depend on the target image, the noise sources involved, and the chosen reconstruction quality metric (dimensionless residuals as we have defined them may not be the most relevant metric for every application).  

A further consideration when choosing $\beta$ for practical applications is that the number of patterns given a significant weight by the NNLS also depends on the value of $\beta$. Smaller $\beta$ values tend to result in many patterns being given similar weights, while larger values of $\beta$ result in only a small number of patterns being given large weights. Recall, in this context, that weights typically correspond to exposure times. Hence, smaller values of $\beta$ require more precise exposure times in practical implementations, while larger values of $\beta$ lead to experiments that are more robust to uncertainty in exposure times.

\begin{figure}[!htb]
    \centering
    \includegraphics[width=\linewidth]{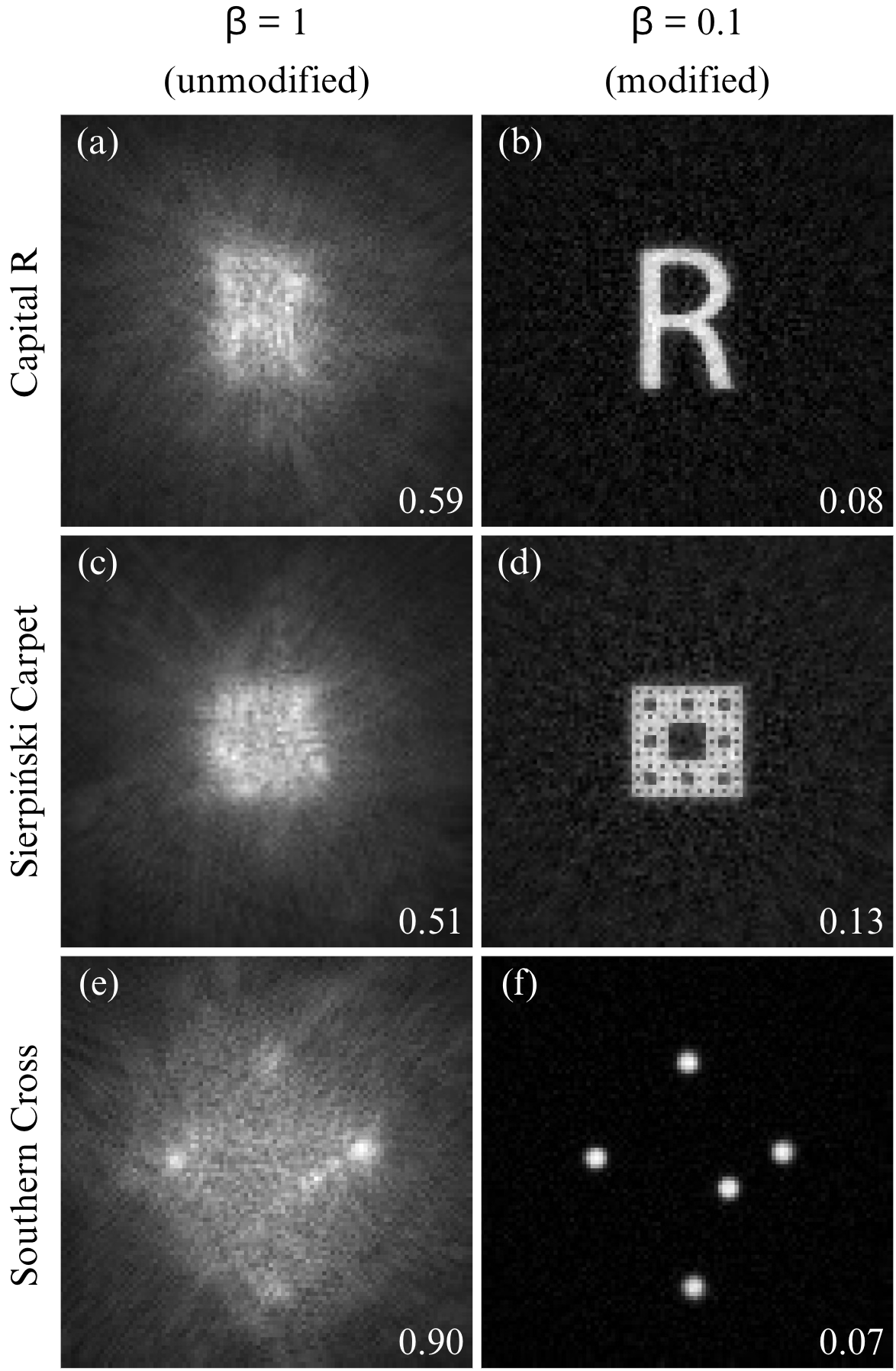}
    \caption{ A comparison between NNLS applied to the original target image (i.e.~$\beta = 1$) and NNLS applied to the modified target image (with $\beta=0.1$), where the AIP has been subtracted from the result. The values in the bottom right corners of each reconstruction are a measure of the residual between the desired image and the reconstruction, minimized with respect to uniform intensity offsets. For a video showing the ghost projection process for panel (d), see the Supplemental Material \cite{SupplementalMaterial}. }
    \label{fig:NNLS vs modified NNLS}
\end{figure}

Figure \ref{fig:NNLS vs modified NNLS} displays the improvement in reconstruction fidelity when various target images are modified according to Eq.~(\ref{eq:beta}), using $\beta =0.1$. The images are constructed using NNLS, and for $\beta = 0.1$, the pedestals are flattened after reconstruction (Sec.~\ref{sec:flattening pedestal} details potential practical methods for flattening the pedestal). Three different $100 \times 100$ original target images were chosen for demonstration: (i) a capital letter ``R'' to test the ghost-projection reconstruction of line-like features (Figs.~\ref{fig:NNLS vs modified NNLS}(a-b)), (ii) the first 3 iterations of the Sierpi\'{n}ski carpet fractal to test fine detail and multiple length scales (Figs.~\ref{fig:NNLS vs modified NNLS}(c-d) and associated video in Supplemental Material \cite{SupplementalMaterial}), and (iii) the Southern Cross stellar constellation to test isolated spots akin to the larger spots in the illuminating patterns (Figs.~\ref{fig:NNLS vs modified NNLS}(e-f)).

Recall that the dimensionless residuals shown in Fig.~\ref{fig:NNLS vs modified NNLS} are defined as the norm squared of the difference divided by the norm squared of the target image. For assessing synthesis quality, this metric is crude in the sense that it is not sensitive to fine detail. For instance, the residual for (c) suggests it is the most faithful of the NNLS reconstructions performed on the original target image, yet it can easily be argued that (a) and (e) capture the majority of the information of their target images, while any details beyond the first iteration of the Sierpi\'{n}ski carpet are lost in (c). 

The reconstructions (b), (d), and (f) in Fig.~\ref{fig:NNLS vs modified NNLS} all show significant improvement, both according to the reduced residual and reconstruction of fine detail. Original target images that are initially closer in intensity profile to the pedestal (e.g.~(a) and (c)) are recreated more successfully, and adding the AIP to the target image has a more pronounced impact for original target images which differ more from the average pattern (e.g.~(e) compared with (f)). 

Regarding fine detail, Fig.~\ref{fig:NNLS vs modified NNLS}(d) shows that when using the modified target image, the high spatial frequency portions of the diffraction catastrophes (single-pixel feature sizes and smaller) allow single-pixel information in the target image to be accurately resolved. 
Moreover, for all target images tested in this study, when NNLS is applied to the modified target image, the additional pedestal (on top of the AIP) is spatially constant, so subtracting the AIP produces a uniform flat background in Figs.~\ref{fig:NNLS vs modified NNLS}(b), (d), and (f). 

If ghost projection is used to encode data to be read digitally, then the pedestal-subtraction process can be conducted digitally after scanning. However, if a flat pedestal is necessary (e.g.~for lithography), then the AIP needs to be subtracted physically (up to a constant shift).  In the latter case, it may be possible to subtract the pedestal during the projection process using an attenuating mask. Alternatively, a suitable intensity distribution may be superimposed afterwards to flatten the pedestal. These ideas are explored in more detail in Sec.~\ref{sec:futureWork}.

\subsection{Influence of Poisson noise}
\label{sect:results_noise}

\begin{figure}[!htb]
    \centering
    \includegraphics[width=\linewidth]{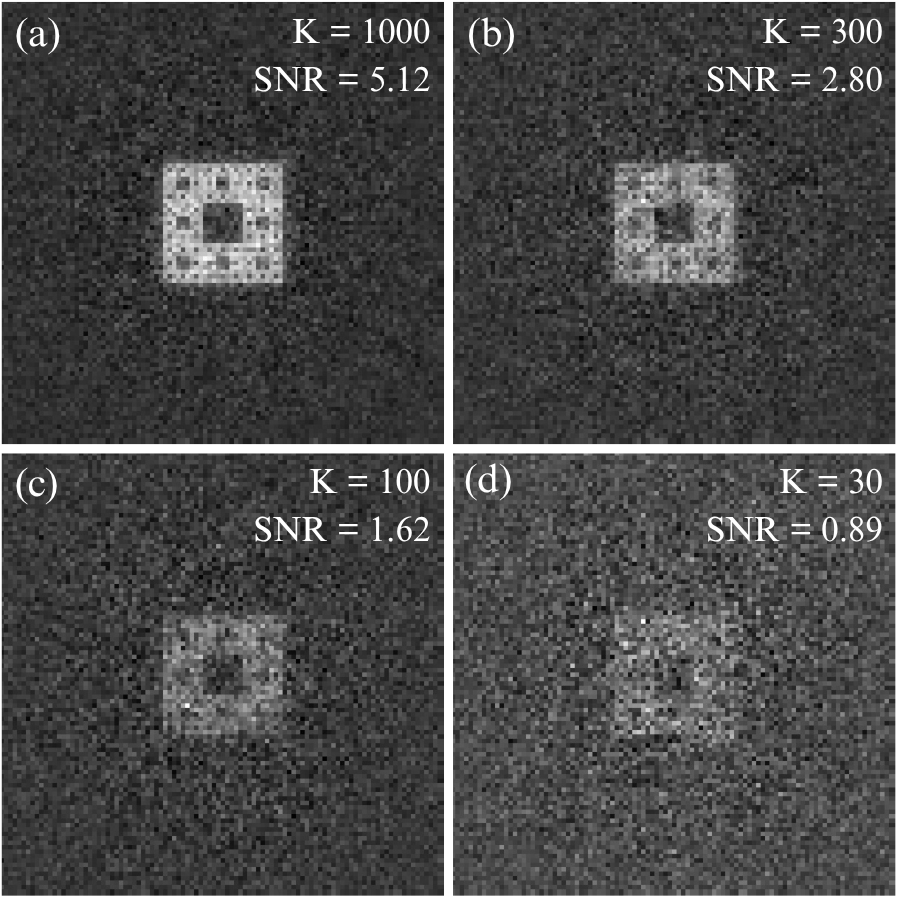}
    \caption{ Simulated ghost projections of a Sierpi\'{n}ski carpet target with $\beta =0.1$ in the presence of Poisson noise, shown for four K values: (a) $K=1000$,  (b) $K=300$, (c) $K=100$, (d) $K=30$. The corresponding SNR values are displayed accordingly. }
    \label{fig:Poisson noise}
\end{figure}

Any practical implementation of our ghost-projection technique would be negatively influenced by photon shot-noise, given by a Poisson distribution, due to quantization of radiant exposure. For the binary target images used in Fig.~\ref{fig:NNLS vs modified NNLS}, $\beta=0.1$ was used. Let $M$ denote the fraction of pixels in a binary target image that are white (with the remaining being black). If a white (i.e., maximum exposure) pixel of the target image receives an average of $K$ more photons than a black (i.e., minimum exposure) pixel of the target image at the same location, Eq.~(\ref{eq:global SNR}) implies
\begin{equation}
\text{SNR} \approx \frac{K \sqrt{M}}{\sqrt{K(1 + \frac{1-\beta}{\beta})}} = \sqrt{\beta KM}. 
\end{equation}
To exceed any desired SNR for these binary images, the projection needs to be exposed for long enough that $K$ satisfies
\begin{equation}
K \gtrapprox \frac{\text{SNR}^2}{\beta M}. 
\end{equation}
Figure~\ref{fig:Poisson noise} shows an example of this for multiple $K$ values, using a Sierpi\'{n}ski carpet target image with $\beta = 0.1$. While we have here focused on the influence of SNR, for lithographic purposes, the required exposure may be determined not only by the SNR but also by the sensitivity of the lithographic substrate.

\section{Discussion}\label{sec:Discussion}

\label{sec:general remarks}

A statistical ensemble, say, of aberrated focal-field diffraction catastrophes, gives an overcomplete non-negative basis\footnote{Here, ``non-negative basis'' refers to a set of non-negative pixellated intensity distributions that form a mathematical basis (modulo a pedestal) for pixellated intensity distributions when linear combinations of the set are formed with non-negative weighting coefficients.  The stated linear combinations form non-negative intensity distributions, but when the pedestal is removed, the resulting distribution may be (and in general will be) both positive and negative. 
As an illustrative example, see Fig.~4(b) in Ref.~\cite{ceddia2023universal}, whose non-negative ghost-projection intensity distribution contains both negative and positive contributions relative to the pedestal.} that may be used to synthesize any desired target intensity distribution.  In this synthesis, the positive weighting coefficients $w_k$ are proportional to the exposure time for each pattern. The preceding two-sentence summary, of the core ``ghost projection'' \cite{paganin2019writing, ceddia2022ghost, ceddia2022ghostII, ceddia2023universal,kingston2025neutron} idea that is explored in the present simulation-based paper, is qualified by the facts that (i) the synthesized target image has a spatial resolution that is governed by the finest features present in the diffraction-catastrophe ensemble, (ii) the target image is augmented by a smoothly-varying additive offset (or ``pedestal''), and (iii) the transverse spatial extent of the synthesized image is limited by the transverse spatial extent of the basis patterns.  Rather than seeking to shape the target field in a direct manner---as would be the case for image synthesis using spatial light modulation, illumination of a single-patterned mask, or raster scanning of a tightly focused beam---our approach is indirect.  In particular,  our method recasts aberration-induced focal-field distortions as an enabling feature, when they can be created in a controllable and reproducible manner, since they generate the basis which underpins our indirect approach to focal-plane image synthesis.  If we loosely speak of the aberration-catastrophe focal-field ensemble as a set of deterministic albeit highly structured ``noise'' maps, then our indirect approach may be spoken of as ``building signals out of noise'' \cite{paganin2019writing} (cf.~Fig.~\ref{fig:VanillaGhostProjection}). 
 Some particular approaches for achieving this means of beam shaping, for a variety of radiation and matter wave fields, are briefly explored in Sec.~\ref{sec:translation to xrays electrons etc}.  Remarks on the advantages of a diffraction-catastrophe basis, together with the more general question of the efficiency of ghost projection using overcomplete basis sets, are given in Sec.~\ref{sec:Efficiecy of random bases}.      

\subsection{Focal-field ghost projection using visible light, x rays, neutrons, and electrons}\label{sec:translation to xrays electrons etc}

Given that the focal-field ghost-projection method is applicable to a variety of radiation and matter wave fields, we now comment on associated differences in possible means for its future experimental implementation.

\begin{figure}[!htb]
    \centering
    \includegraphics[width=\linewidth]{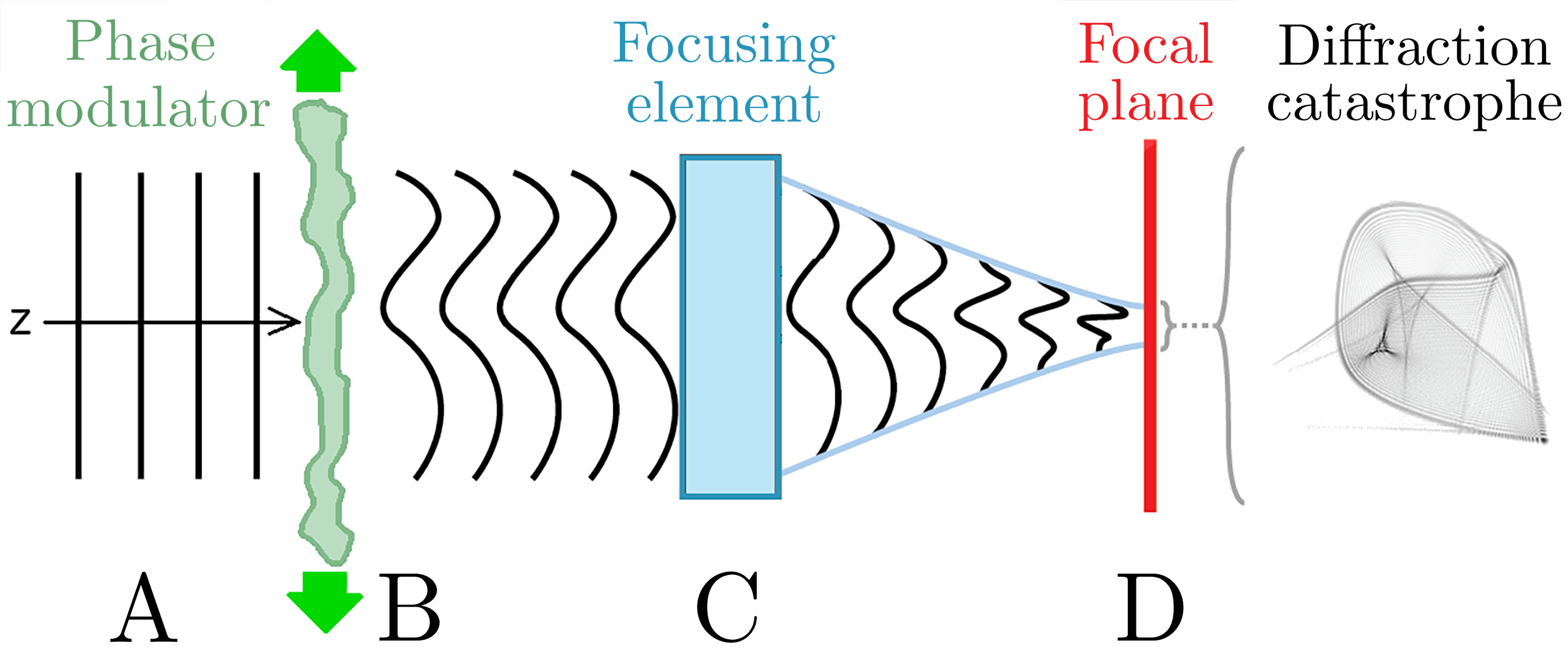}
    \caption{Coherent plane waves $A$ illuminate a smooth transparent spatially-random screen $B$, that may be reproducibly and accurately displaced in two directions transverse to the optical axis $z$. A focusing element $C$ then creates a diffraction catastrophe in the associated focal plane $D$.  A suitable linear superposition of such focused fields yields a summed intensity distribution over the focal plane which approximates any desired target pattern, up to a limiting spatial resolution and additive pedestal.}
    \label{fig:SpatiallyRandomScreen}
\end{figure}

\begin{enumerate}[(i)]

    \item For visible light, the phase modulator in Fig.~\ref{fig:Generating diffraction catastrophes} can be a spatial light modulator (SLM) \cite{SLMpaper}.  Alternatively, as shown in Fig.~\ref{fig:SpatiallyRandomScreen}, the phase modulator can comprise a smoothly-varying random-thickness transparent screen $B$ on a transverse translation stage, with the focusing optic $C$ being a glass lens.

    \item In implementations for hard x rays, the phase modulator $B$ in Fig.~\ref{fig:SpatiallyRandomScreen} could consist of a transversely-displaced spatially-random amorphous screen that is weakly absorbing at such photon energies (e.g., a variable-thickness amorphous carbon film \cite{Falch2019}, a beryllium window \cite{SNIGIREV1996634} with smooth random variations in projected thickness, or a Kapton film \cite{SuzukiUchida1995} with variable projected thickness).  An alternative means for phase modulation could employ a deformable mirror operating at grazing angles of incidence \cite{XRayDeformableMirror2024}. The focusing optic $C$ may be a Fresnel zone plate \cite{Kirz1974}, compound refractive lens \cite{SnigirevCRL,Tomie2010}, or Kirkpatrick--Baez mirror \cite{Kirkpatrick1948Formation1948}. To obtain a sufficient degree of coherence, in all of the x-ray scenarios described above, energy-filtered x-ray synchrotron radiation \cite{HofmannSynchrotronBook} could be employed.

    \item Neutron focal-field ghost projection could also employ weakly-absorbing spatially-random transversely-displaced screens $B$ as a phase modulator, with suitable neutron-transparent materials including lead and aluminum \cite{NeutronImagingAndItsApplications}.  For the focusing optic $C$, compound refractive lenses \cite{NeutronCRL1,NeutronCRL2} or Fresnel zone plates \cite{KearneyKleinOpatNeutronZonePlate} could be used. 

    \item In a modern aberration-corrected transmission electron microscope (TEM) \cite{AberrationCorrectedTEM==2023Review}, phase modulation could be achieved by tuning magnetic-lens aberrations such as coma, astigmatism, and various orders of spherical aberration; the focusing lens in this case is a suitably shaped magnetic field \cite{ReimerKohlBook}. As an alternative, the scenario in Fig.~\ref{fig:SpatiallyRandomScreen} could be employed, with $B$ being an electron-transparent amorphous carbon film \cite{SpenceTEMBook} and $C$ being the TEM magnetic-lens system.
    
\end{enumerate}

\subsection{Advantages of a diffraction-catastrophe basis}\label{sec:Efficiecy of random bases}

Diffraction-catastrophe bases naturally arise from a beam-focusing configuration. First of all, they are straightforward to produce: any sufficiently strong perturbation of the beam's phase will cause a highly-structured diffraction catastrophe at the focal plane, even if the phase perturbation itself is smooth and undetailed. Also, the patterns have been observed to contain features at a range of length scales, which is advantageous when constructing arbitrary images. It is difficult to produce some other basis (such as a binary speckle pattern) with such a setup. Physical attenuating masks are plagued by distorting diffraction effects at small scales (i.e.~when the features are comparable to the wavelength). It is also impractical to reverse-engineer the necessary phase profile to project an arbitrary pattern upon focusing; not only is it a mathematically challenging task, but an arbitrary pattern will require an arbitrarily complicated phase profile, which may be impossible to accurately construct with the phase modulator. The diffraction catastrophe bases discussed in this paper require simple phase profiles by design, as the phase profiles are made using only the first 21 Zernike polynomials.

We now turn to a second point of advantage of the diffraction-catastrophe basis, which relates to its overcomplete nature.  Complete orthogonal bases are often considered to be preferable to non-orthogonal bases, in a range of problems pertaining to synthesis or decomposition in the context of optical physics. This optimality of orthogonal complete bases may be challenged by nonorthogonal overcomplete \cite{MandelWolf} bases, {\em when seeking to synthesize one particular arbitrary function}. Indeed, one can choose a very small subset of a highly overcomplete basis, with the particular subset being chosen such that the representation of the ghost projection is sparse (i.e., the number of members in the subset is small).  Thus stated, the ghost-projection method may be optimized via a reversed form of compressive sensing \cite{katz2009compressive}. This reversed form of compressive sensing is implicitly performed in our NNLS approach to ghost projection, as evidenced by the large fraction of weights, $w_k$, that are set to zero by this approach (see Fig.~\ref{fig:Normal reconstructions}(d)).  Indeed, since the target image occupies a finite area $\mathcal{A}$ in the lens' focal plane, there is a finite number of degrees of freedom
\begin{equation}
\frak{N}\approx\mathcal{A}/\ell^2 
\end{equation}
associated with the target image, where $\ell$ is the spatial resolution of the target image.  Considered as an overcomplete basis, the focal-field diffraction-catastrophe ensemble $\frak{E}$ has infinitely many members, from which a subset $\frak{S}$ having on the order of $\frak{N}$ (or fewer) members can be chosen, in order to synthesize a specified target image.  Given that one has total prior knowledge of the target one wishes to synthesize, there is significant freedom in choosing $\frak{S}$ for the purposes of synthesizing one particular target image; this choice can be made subject to the previously implied constraint that the number of members in $\frak{S}$ be on the order of $\frak{N}$ or less, although the method is sufficiently flexible for other constraints or criteria to be employed.

\section{Practical considerations and future work}\label{sec:PracticalConsiderationsAndFuturework}

As demonstrated for the example in Sec.~\ref{sec:results_modifying_target}, adding the average pattern before applying NNLS increases the number of patterns that significantly contribute to the synthesized image; approximately 3\% of the patterns (as opposed to 0.4\%) are assigned a weight greater than 1\% of the maximum weight. However, it has been observed that as a percentage, the number of patterns that contribute decreases as the total number of patterns available to the NNLS algorithm increases, and in some cases the absolute number of contributing patterns may decrease when particularly suitable patterns are introduced.

Suppose that removing patterns with low weights (e.g.~those below 1\% of the maximum weight) has negligible effect on the synthesized image quality. The NNLS reconstructions assign large exposure times (weights) to a small number of patterns, so when the negligible-weight patterns are omitted, each illuminating pattern is exposed for longer and the uncertainty introduced by the shutter speed is a smaller fraction of the exposure time for each pattern. This improves the resilience of the reconstruction to exposure-time noise. Additionally, if a non-negligible amount of time is required to adjust the phase modulation component between each illuminating pattern, then using fewer patterns decreases the required total projection time. It may be possible to reduce this adjustment time by ordering the illuminating patterns so that the phase changes minimally between consecutive patterns (i.e.~the Zernike polynomial coefficients change minimally). Depending on the density of contributing illuminating patterns, there may need to be no beam downtime, and the illuminating pattern may be allowed to evolve continuously.

Note that by generating more illuminating patterns for the NNLS to use, computation time can be traded for more efficient and accurate image synthesis. This is especially valuable if the target image is one that will be mass produced, as the set of weights only needs to be determined once.

\label{sec:flattening pedestal}

Some applications, such as lithography, will require a spatially-uniform (``flat'') pedestal.  We present three possible avenues for pedestal flattening: (i) using an appropriate attenuating mask during projection, (ii) superposing an inverted pedestal ``after the fact'', or (iii) sidestepping the issue by restricting the ghost projection field of view to be sufficiently small:

\begin{enumerate}[(i)]
    \item Assuming the pedestal to be independent of the target image, it should be possible to produce a single attenuating mask to insert between the lens and the focal plane such that the resulting projection has a flat pedestal. Using a material with attenuation coefficient $\mu$, the transmission through the mask is $e^{-\mu T_1(\mathbf{r})}$, where $T_1$ is the thickness of the mask along the direction of the optical axis which is perpendicular to the transverse coordinate $\mathbf{r}$. Let $\eta < \min\langle{I}(\mathbf{r})\rangle$ be the intensity of the desired uniform pedestal, i.e.~we are looking for a thickness function, $T_1$, such that $\langle I(\mathbf{r})\rangle e^{-\mu T_1(\mathbf{r})} = \eta$. Rearranging, we get the necessary thickness of the mask
    \begin{equation}
    T_1(\mathbf{r}) = \frac{1}{\mu} \ln\left(\frac{\langle{I}(\mathbf{r})\rangle}{\eta}\right).  
    \end{equation}
    This method requires a few additional considerations: the refractive properties of the mask will distort the diffraction catastrophes, so they need to be re-calculated or measured before the NNLS weights are generated. This may require high precision in the machining and positioning of the attenuating mask, as the fine details in the diffraction patterns will be highly sensitive to the placement and mask profile.
    The target image intensity must be boosted (multiplied by $e^{\mu T_1(\mathbf{r})}$) to compensate for the attenuation, ensuring that the desired projected image is attained only after passing through the attenuating mask. 
    \item If the precision necessary for (i) is not achievable, it is possible to flatten the pedestal retrospectively, by projecting the negative pedestal (up to a shift, as negative exposure is not possible). Precisely, one would superpose
    \begin{equation}
        \mathcal{P}^*(\mathbf{r}) = C - \langle I(\mathbf{r})\rangle
    \end{equation}
    where $C > \max\langle I(\mathbf{r})\rangle $ is the value of the resulting flat pedestal. This can also be achieved using an attenuating mask. For a material with attenuation coefficient $\mu$ and transverse thickness function $T_2$, we require $\langle{I}(\mathbf{r})\rangle + Ae^{-\mu T_2(\mathbf{r})} = C$, where $A$ is the total uniform exposure incident upon the mask. The required thickness function is 
    \begin{equation}
        T_2(\mathbf{r}) = \frac{1}{\mu}\ln\left( \frac{A}{C-\langle I(\mathbf{r})\rangle} \right).
    \end{equation}
    The advantage of this method is that as the pedestal flattening is separate from the image construction, the refractive properties of the attenuating mask will not warp the interference patterns, and so the precision in machining and placement of the mask does not need to be nearly as high. The disadvantage is that the resulting flat pedestal, $C$, is larger than the flat pedestal, $\eta$, in method (i), since $C > \max \langle I (\mathbf{r})\rangle > \min \langle I(\mathbf{r})\rangle > \eta$. This means that that the signal-to-noise ratio is preferable for method (i).
    \item Pedestal flattening can also be achieved in a simple manner by confining the ghost projection to a region near the pedestal's ``center of mass'' that has a diameter which is small compared to the width of the pedestal. In other words, the field of view can be cropped to a region over which the pedestal is approximately constant. This comes at the price of a lower resolution image.
\end{enumerate}

\label{sec:futureWork}

A number of avenues for future work remain. Experimental realization of the simulated work in this paper is necessary to establish the viability of using diffraction catastrophes for ghost-projection lithography. For instance, to compete with electron-beam lithography that has no pedestal, it must be demonstrated that the sub-Airy-disk detail within the diffraction catastrophes can be physically recreated to sufficient precision, as otherwise the fine details that make this method effective are not available for use. Furthermore, considerations such as the numerical aperture of the lens (or equivalent for other focusing systems) need to be incorporated.\footnote{It is a standard optical-physics result that the resolution associated with a lens-based imaging system is on the order of the imaging-radiation wavelength $\lambda$ divided by the numerical aperture $N_A$ (see e.g.~p.~466 of Ref.~\cite{born2013principles} or p.~144 of Ref.~\cite{Messiah}).  It would be interesting to seek a corresponding result for ghost projection via focal-field diffraction catastrophes, namely the relationship between $N_A$ and ghost-projection spatial resolution $s$.  We conjecture that $s\approx F\lambda/N_A$, where $F$ is a dimensionless multiplicative factor on the order of unity, but further work is needed to investigate if this suggestion is correct.} Also, the effect of imprecision in the phase modulation process and inaccuracy in the illuminating pattern exposure times requires investigation to determine usable parameters for practical purposes.

Computationally, areas of further exploration include ascertaining the relationship between the number of patterns available to the NNLS algorithm, and the number of patterns given significant weight. This relationship will likely depend on the desired target image.  The threshold for considering weights to be significant may also be examined to evaluate the impact of discarding patterns below this threshold. 

Besides the mentioned applications in dynamic on-demand beam shaping, aberration correction, and lithography, a further potential application is tomographic volumetric additive manufacturing, which is an alternative technique to conventional layer-by-layer additive manufacturing (standard 3D printing) that involves solidifying a photopolymer by irradiating it from multiple angles with dynamic light patterns \cite{Kelly2019,tomographic-volume-additive-manufacturing,Toombs2022}. This would utilize the three-dimensional profile of the diffraction catastrophes, requiring analysis of how the diffraction patterns evolve with distance along the beam axis. As demonstrated in e.g.~Ref.~\cite{tomographic-volume-additive-manufacturing}, tomographic volumetric additive manufacturing already achieves smaller printing scales than layer-by-layer manufacturing, and if the material properties of the photopolymer are not the limiting factor, diffraction-catastrophe ghost projection may allow further miniaturization.

\section{Conclusion}\label{sec:Conclusion}

 Ghost projection creates a desired distribution of radiant exposure by cumulatively exposing a fixed set of illuminating patterns. These basis-function patterns may have a random character, although the method may also employ nonrandom patterns. The technique developed in this study uses highly-structured diffraction catastrophes as basis functions, produced by focusing a coherent beam whose phase has been continuously modulated in a random manner. The approach, which may be viewed as a reversed form of classical computational ghost imaging, has been demonstrated via simulation. Possible future applications of focal-field ghost projection include dynamic on-demand beam shaping of focused fields, an indirect form of aberration correction, and lithography.

\begin{acknowledgments}
AMK and DMP thank the Australian Research Council (ARC) for funding through the Discovery Project: DP210101312. AMK thanks the ARC and Industry partners funding the Industrial Transformation and Training  Centre for Multiscale 3D Imaging, Modelling, and Manufacturing: IC180100008. This research utilized resources and services from the National Computational Infrastructure (NCI) facility, which is supported by the Australian Government. We acknowledge useful discussions with Christian~Dwyer, Wilfred~K.~Fullagar, Kieran Larkin, Tanya M.~Monro, and  Timothy C.~Petersen.

\end{acknowledgments}


\bibliography{references}

\end{document}